%Paper: hep-th/9212147
%From: khare@iopb.ernet.in
%Date: Wed Dec 23 11:28:33 1992

\magnification=\magstep1
\hsize = 32 pc
\vsize = 42 pc
\baselineskip = 24 true pt
\centerline {\bf New Shape Invariant Potentials in Supersymmetric}
\centerline {\bf Quantum Mechanics}
\vskip 0.5 true cm
\centerline {\bf Avinash Khare and Uday P. Sukhatme$^*$}
\centerline {Institute of Physics, Sachivalaya Marg,}
\centerline {Bhubaneswar 751005, India}
\vskip 0.5 true cm
\noindent {\bf Abstract:} Quantum mechanical potentials satisfying
the property of shape invariance are well known to be algebraically
solvable. Using a scaling ansatz for the change of parameters,
we obtain a large class of new shape invariant potentials
which are reflectionless and possess an infinite number of
bound states. They can be viewed as q-deformations of the single
soliton solution corresponding to the Rosen-Morse potential.
Explicit expressions for energy eigenvalues, eigenfunctions and
transmission coefficients are given. Included in our potentials as a
special case is the self-similar potential recently discussed by
Shabat and Spiridonov.
\vskip 3.0 true cm
\item {*} Permanent Address: Department of Physics, University of
Illinois at Chicago.
\vfill
\eject
In recent years, supersymmetric quantum mechanics [1] has yielded
many interesting results. Some time ago, Gendenshtein pointed out
that supersymmetric partner potentials satisfying the property of
shape invariance and unbroken supersymmetry are exactly solvable [2].
The shape invariance condition is
$$V_+(x,a_0)=V_-(x,a_1)+R(a_0),\eqno{(1)}$$
where $a_0$ is a set of parameters and $a_1=f(a_0)$ is an arbitrary
function describing the change of parameters.
The common x-dependence in $V_-$ and $V_+$ allows full
determination of energy eigenvalues [2], eigenfunctions [3] and
scattering matrices [4] algebraically. One finds $(\hbar = 2m = 1)$
$$E^{(-)}_n(a_0)=\sum^{n-1}_{k=0}
R(a_k),\qquad E^{(-)}_0(a_0)=0,\eqno{(2)}$$
$$\psi^{(-)}_n(x,a_0)=\Bigg [-{d\over
dx}+W(x,a_0)\Bigg ]\psi^{(-)}_{n-1}(x,a_1),
\quad\psi^{(-)}_0(x,a_0)\quad\propto
\quad e^{-\int^x W(y,a_0)dy}\eqno{(3)}$$
where the superpotential $W(x,a_0)$ is related to $V_{\pm}(x,a_0)$ by
$$V_{\pm}(x,a_0) = W^2(x,a_0)\pm W^{\prime}(x,a_0).\eqno{(4)}$$
In terms of W, the shape invariance condition  reads
$$W^2(x,a_0)+W^{\prime}(x,a_0)
=W^2(x,a_1)-W^{\prime}(x,a_1)+R(a_0).\eqno{(5)}$$
It is still a challenging open problem
to identify and  classify the solutions to
eq.(5). Certain solutions to the shape invariance
condition are known [5]. (They include the
harmonic oscillator, Coulomb, Morse, Eckart and
Poschl-Teller potentials).  In all these cases,
it turns out that $a_1$ and $a_0$ are
related by a translation $(a_1=a_0+\alpha)$. Careful searches with this
ansatz have failed to yield any additional shape invariant potentials
[6]. Indeed it has been suggested [7] that there are no other shape
invariant potentials. Although a rigorous proof has never been
presented, no counter examples have so far been found either.

In this letter, we consider solutions of eq.(5)
resulting from a new scaling ansatz
$$a_1 =q a_0,\eqno{(6)}$$
where $0<q<1$. This choice was motivated by the recent interest in
q-deformed Lie algebras. It enables us to find a large class
of new shape invariant potentials all of which are reflectionless and
possess an infinite number of bound states. As a special case, our
approach includes the self-similar potential studied by Shabat [8]
and Spiridonov [9].

Consider an expansion of the superpotential of
the form
$$W(x,a_0)=\sum^{\infty}_{j=0} g_j(x) a^j_0.\eqno{(7)}$$
Substituting into eq.(5), writing $R(a_0)$ in the form
$$R(a_0) = \sum^{\infty}_{j=0} R_ja^j_0,\eqno{(8)}$$
and equating powers of $a_0$ yields
$$g'_0(x) = R_0, \quad g_0(x) = R_0 x+C_0,\eqno{(9)}$$
$$g'_n(x)+2d_n g_0(x)g_n(x)
= r_n-d_n \sum^{n-1}_{j=1} g_j(x)g_{n-j}(x),\eqno{(10)}$$
where
$$R_n\equiv (1-q^n)r_n,\quad d_n\equiv
{1-q^n\over 1+q^n}\quad (n = 1,2,3...).\eqno{(11)}$$
This set of linear differential equations is easily solvable is
succession yielding a general solution of eq.(5). Note that the limit
$q\rightarrow 0$ is particularly simple yielding the one-soliton
Rosen-Morse potential of the form $W=\gamma tanh \gamma x$. Thus our
results can be regarded as multiparameter deformations of this
potential corresponding to different choices of $R_n$. For
simplicity, in this letter we shall confine our attention to the
special case $g_0(x)=0$ (i.e. $R_0=C_0=0)$ while the more general
case will be discussed elsewhere [10].

For $g_0=0$, the solution is
$$g_n(x)=d_n\int dx\Bigg [r_n-\sum^{n-1}_{j=1}
g_j(x)g_{n-j}(x)\Bigg ].\eqno{(12)}$$

For the simplest case of $r_n=0, n\geq 2$ we obtain the superpotential W
as given by Shabat [8] and Spiridonov [9] provided we choose $d_1r_1 a_0
=\gamma^2$ and replace q by $q^2$. This shows that the
self-similarity condition of these authors is in fact a special case
of the shape invariance condition (5). This comment is also true in
case any one $r_n$ (say $r_j$) is taken to be nonzero and $q^j$ is
replaced by $q^2$.

Let us now consider a somewhat more general case when $r_n=0, n\geq
3$. Using eq.(12) we can readily calculate all $g_n(x)$. The first
three are
$$g_1(x) = d_1r_1x,\quad g_2(x)= d_2r_2x -{1\over 3} d^2_1r^2_1d_2 x^3,$$
$$g_3(x)=-{2\over 3} d_1r_1d_2r_2 d_3 x^3 + {2\over 15}
d^3_1r^3_1d_2d_3  x^5.\eqno{(13)}$$
Note that W(x) contains only odd powers of x.
This makes the potential $V_-(x)$ symmetric in x and also
guarantees the situation of unbroken supersymmetry. The energy
eigenvalues follow immediately from eqs.(2) and (8)
(n=0,1,2.....$\infty; 0 <q<1)$
$$E^{(-)}_n(a_0) = d_1r_1 a_0{(1-q^n)\over (1-q)}
+d_2r_2 a^2_0{(1-q^{2n})\over (1-q^2)}.\eqno{(14)}$$
Note that the energy eigenvalues, the superpotential
W and hence the eigenfunctions only
depend on the two combinations of parameters $\gamma^2_1\equiv
d_1r_1 a_0$ and $\gamma^2_2\equiv d_2r_2 a^2_0$. The unnormalized
ground state wave-function is
$$\psi^{(-)}_0(x,a_0)=exp\Bigg [-{x^2\over
2}(\gamma^2_1+\gamma^2_2)+{x^4\over 12}(d_2\gamma^4_1+2
d_3\gamma^2_1\gamma^2_2+d_4\gamma^4_2)+0(x^6)\Bigg ].\eqno{(15)}$$
The excited state wave-functions can be recursively calculated by
using eq.(3) with $a_1=q a_0$.

The transmission coefficient of two symmetric partner potentials are
related by [11]
$$T_-(k,a_0)=\Bigg [{ik-W(\infty,a_0)\over
ik+W(\infty,a_0)}\Bigg ] T_+(k,a_0),\eqno{(16)}$$
where $k=[E-W^2(\infty,a_0]^{1/2}$. For a shape invariant potential
$$T_+(k,a_0) = T_-(k,a_1).\eqno{(17)}$$
Repeated application of eqs.(16) and (17) gives
$$T_-(k,a_0)=\left [{ik-W(\infty,a_0)\over ik+W(\infty,a_0)}\right
]\left [{ik-W(\infty,a_1)\over ik+W(\infty,a_1)}\right ]...\left
[{ik-W(\infty,a_{n-1})\over ik+W(\infty,a_{n-1})}\right ]T_-(k,a_n),
\eqno{(18)}$$
where
$$W(\infty,a_j)=\sqrt{E^{(-)}_{\infty}-E^{(-)}_j}.\eqno{(19)}$$
As $n\rightarrow\infty$, since $a_n=q^na_0$ and we have taken
$g_0(x)=0$, one gets $W(x,a_n)\rightarrow 0$. This corresponds to a
free particle for which the transmission coefficient is unity. Thus,
for the potential $V_-(x,a_0)$, the
reflection coefficient vanishes and the transmission coefficient is
$$T_-(k,a_0)= \Pi^{\infty}_{j=0}\quad \left [ {ik-W(\infty,a_j)\over
ik+W(\infty,a_j)}\right ].\eqno{(20)}$$
Clearly, $\mid T\mid^2 = 1$ and the poles of $T_-$ correspond
to the energy eigenvalues of eq.(14). Note that one does not get
reflectionless potentials for  the case $g_0(x)\not = 0$.
This will be further
discussed in ref.[10].

The above discussion, keeping only $r_1,r_2\not =0,$ can readily be
generalized to an arbitrary number of nonzero $r_j$. The energy
eigenvalues for this case are given by $(\gamma^2_j = d_jr_ja^j_0)$
$$E^{(-)}_n(a_0)=\sum_j\gamma^2_j
({1-q^{jn}\over 1-q^j}),\quad n = 0,1,2,....\eqno{(21)}$$
All of these potentials are also reflectionless with $T_-$ as given by
eqs.(19) to (21). One expects that these symmetric reflectionless
potentials can also be derived using previously developed methods
[12] and the spectrum given in eq.(21).

In eqs.(7) and (8) we have only kept positive powers of $a_0$. If
instead we had only kept negative powers of $a_0$, then the spectrum
would be similar except that one has
to choose the deformation parameter $q>1$.
However, a mixture of positive and
negative powers of $a_0$ is not allowed in general since neither $q <
1$ nor $q
> 1$ will give an acceptable spectrum. For the enlarged class of
shape invariant potentials discussed in this letter, it is clear [3]
that the lowest order supersymmetric WKB approximation [13] will
yield the exact spectrum.

We conclude with two brief remarks on extensions of the work
described in this letter. We have been able to construct new shape
invariant potentials which are q-deformations of the potentials
corresponding to multi-soliton systems [10]. Also, it is possible to
show [10] that with the choice $g_0(x) \not = 0,$ one gets
q-deformations of the one dimensional harmonic oscillator potential.
\vskip 0.5 true cm

One of us [U.S.] would like to thank the Council of Scientific and
Industrial Research and the  United Nations Development
Programme for support under their TOKTEN scheme and the kind
hospitality of the Institute of Physics, Bhubaneswar where this work
was done. This work was supported in part by the U.S. Department of Energy.
\vfill
\eject
\noindent {\bf References}
\item {[1]} E.Witten, Nucl. Phys. {\bf B185}, 513 (1981); F. Cooper
and B. Freedman, Ann. Phys. (N.Y) {\bf 146}, 262 (1983).
\item {[2]} L.Gendenshtein, JETP Lett. {\bf 38}, 356 (1983).
\item {[3]} R. Dutt, A. Khare and U.P.Sukhatme, Phys. Lett. {\bf
181B}, 295 (1986); J.Dabrowska, A.Khare and U.P.Sukhatme, J. Phys. A:
Math. Gen. {\bf 21}, L195 (1988).
\item {[4]} A.Khare and U.P.Sukhatme, J. Phys. A: Math. Gen. {\bf
21}, L501 (1988).
\item {[5]} R.Dutt, A.Khare and U.P.Sukhatme, Am. J. Phys. {\bf 56},
163 (1988); L.Infeld and T.Hull, Rev. Mod. Phys. {\bf 23}, 21 (1951).
\item {[6]} F. Cooper, J.N.Ginocchio and A.Khare, Phys. Rev. {\bf
D36}, 2458 (1987).
\item {[7]} D.T.Barclay and C.J.Maxwell, Phys. Lett. {\bf A157}, 357 (1991).
\item {[8]} A. Shabat, Inverse Prob. {\bf 8}, 303 (1992).
\item {[9]} V.Spiridonov, Phys. Rev. Lett. {\bf 69}, 398 (1992).
\item {[10]} R. Dutt, A.Gangopadhyaya, A.Khare and U.P.Sukhatme, to be
published.
\item {[11]} R.Akhoury and A.Comtet, Nucl. Phys. {\bf B246}, 253
(1984); C.V.Sukumar, J. Phys. A: Math. Gen. {\bf 19}, 2297 (1986).
\item {[12]} W. Kwong, H.Riggs, J.L.Rosner and H.B.Thacker, Phys.
Rev. {\bf D39}, 1242 (1989) and references therein.
\item {[13]} A.Comtet, A.D.Bandrauk and D.K.Campbell, Phys. Lett.
{\bf B150}, 159 (1985); A.Khare, Phys. Lett. {\bf B161}, 131 (1985).
\vfill
\eject
\end